\documentclass[twocolumn,twoside]{IEEEtran}

\usepackage{etex}
\reserveinserts{28}

\ifCLASSINFOpdf
  \usepackage[pdftex]{graphicx}
\else
\fi

\usepackage[colorlinks]{hyperref}
\usepackage{multirow}
\usepackage{times}
\usepackage{amsmath,amssymb}
\usepackage{algorithm}
\usepackage{epstopdf}
\usepackage{subfigure}
\usepackage[dvipsnames]{xcolor}
\usepackage{algorithm}
\usepackage{algorithmic}
\usepackage{cite}
\usepackage{fancyvrb}

\usepackage{bbm}

{\begin{list}               
    {$\bullet$ \hfill}{
        \setlength{\leftmargin}{\parindent}
        \setlength{\parsep}{0.04\baselineskip}
        \setlength{\itemsep}{0.5\parsep}
        \setlength{\labelwidth}{\leftmargin}
        \setlength{\labelsep}{0em}}
    }
{\end{list}}

\providecommand{\eref}[1]{\eqref{#1}}  
\providecommand{\cref}[1]{Chapter~\ref{#1}}

\providecommand{\fref}[1]{Figure~\ref{#1}}

\providecommand{\R}{\ensuremath{\mathbb{R}}}

\providecommand{\E}{\ensuremath{\mathbb{E}}}

\renewcommand{\vec}[1]{\ensuremath{\boldsymbol{#1}}}
\providecommand{\mat}[1]{\ensuremath{\boldsymbol{#1}}}

\providecommand{\calB}{\mathcal{B}}

\providecommand{\calF}{\mathcal{F}}

\providecommand{\calH}{\mathcal{H}}

\providecommand{\calN}{\mathcal{N}}

\providecommand{\calT}{\mathcal{T}}

\providecommand{\mB}{\mathbf{B}}

\providecommand{\mH}{\mathbf{H}}

\providecommand{\mT}{\mathbf{T}}

\providecommand{\vt}{\mathbf{t}}
\providecommand{\vu}{\mathbf{u}}
\providecommand{\vv}{\mathbf{v}}

\providecommand{\vx}{\mathbf{x}}

\providecommand{\mSigma}{\mat{\Sigma}}

\providecommand{\vrho}{\vec{\rho}}

\title{Tilt-then-Blur or Blur-then-Tilt? \\ Clarifying the Atmospheric Turbulence Model}
\author{Stanley~H.~Chan,~\IEEEmembership{Senior Member,~IEEE}
\thanks{The author is with the School of Electrical and Computer Engineering, Purdue University, West Lafayette, IN 47907, USA. Email: \texttt{stanchan@purdue.edu}.}
\thanks{The work is supported in part by the Intelligence Advanced Research Projects Activity (IARPA) under Contract No. 2022-21102100004, and in part by the National Science Foundation under the grants CCSS-2030570 and IIS-2133032. The views and conclusions contained herein are those of the authors and should not be interpreted as necessarily representing the official policies, either expressed or implied, of IARPA, or the U.S. Government. The U.S. Government is authorized to reproduce and distribute reprints for governmental purposes notwithstanding any copyright annotation therein.}
}

\graphicspath{{pix/}}

\begin{document}
\maketitle

\begin{abstract}
Imaging at a long distance often requires advanced image restoration algorithms to compensate for the distortions caused by atmospheric turbulence. However, unlike many standard restoration problems such as deconvolution, the forward image formation model of the atmospheric turbulence does not have a simple expression. Thanks to the Zernike representation of the phase, one can show that the forward model is a combination of tilt (pixel shifting due to the linear phase terms) and blur (image smoothing due to the high order aberrations).

Confusions then arise between the ordering of the two operators. Should the model be tilt-then-blur, or blur-then-tilt? Some papers in the literature say that the model is tilt-then-blur, whereas more papers say that it is blur-then-tilt. This paper clarifies the differences between the two and discusses why the tilt-then-blur is the correct model. Recommendations are given to the research community.
\end{abstract}

\begin{keywords}
Atmospheric turbulence, image restoration, forward model, simulation, convolution, signal processing
\end{keywords}

\section{Introduction}
Over the past decade, there is a significant growth of image processing research focusing on mitigating atmospheric turbulence effects present in images and videos taken by long-range cameras \cite{Anantrasirichai_2013_TIP, Mao_2020_TCI, Li_2021_ICCV, Jin_2021_NatureMI, Feng_2022_TurbuGAN}. Since image restoration is an inverse problem, knowing the forward image formation model is necessary to formulate the restoration problem and derive the optimization algorithm. However, unlike many restoration problems such as deconvolution, atmospheric turbulence does not have a simple equation that can describe how images are distorted. In fact, optics textbooks tell us that the turbulent effect is caused by the changing index of refraction along the optical path \cite{Goodman_2005_FourierOptics,Goodman_2015_StatisticalOptics,Roggemann_1996_Book,Schmidt_2010_Book}. The index of refraction is a stochastic process, and the distortions are realized by how the phase of the wave is perturbed \cite{Tatarski_1967_Book, Fried_1965_Statistics, Fried_1966_Optical}.

From an image processing perspective, we all understand the importance of the forward model but we also realize the difficulty of using wave propagation theory. Thus, in the image processing literature, we often see the so-called ``tilt + blur'' model \cite{Shimizu_2008_CVPR, Leonard_2012_SPIE, Schwarzman_2017_ICCP}. The argument is that the phase distortions will cause the pixels to \emph{shift} (thus the ``tilt'') and the high order aberrations will cause the image to look \emph{smoothed} (thus the ``blur''). Yet, when the two operators are present, the functional composition requires us to specify the \emph{order}. Shall we tilt the image first and then add the blur, or shall we blur the image first and then add the tilt? Or, perhaps it does not matter?

The confusion between the tilt-then-blur and the blur-then-tilt is not a light one. To the best of the author's knowledge, at least in the literature of image processing, both models have been used. For the tilt-then-blur model, one of the earlier papers is by Shimizu et al. published in CVPR 2008 \cite{Shimizu_2008_CVPR}. When they solved the inverse problem, the optimization is implemented for the tilt-then-blur model. Another highly referenced paper by Zhu and Milanfar, published in T-PAMI in 2013, also used the tilt-then-blur model \cite{Milanfar_2013_TPAMI}. On the other side of the spectrum, there is a consistent emphasis of the blur-then-tilt model. One of the earlier work is by Mao and Gilles in 2012 \cite{Gilles_2012_Inverse}, where they modeled turbulence using blur-then-tilt. Mao and Gilles credited the model to Frakes et al. \cite{Frakes_2001_ICASSP} and Gepshtein et al. \cite{Gepshtein_2004_EUSIP}, although neither of Frakes et al. nor Gepshtein et al. had actually discussed the forward model. Subsequent work of Lou et al. \cite{Lou_2013_Inverse} followed the blur-then-tilt model, and this tradition continues until today where many of the latest deep learning based approaches also use the blur-then-tilt model, for example, Lau et al. \cite{Lau_2021_ATFaceGAN}, Nair and Patel \cite{Nair_2021_ICIP}, Yasarla and Patel \cite{Yasarla_2022_CNN}, and Lau and Lui \cite{Lau_2017_Quasiconformal, Lau_2021_Subsampled}.

The goal of this paper is to clarify the difference between the tilt-then-blur model and the blur-then-tilt model. There are three main conclusions:
\begin{itemize}
\item The blur-then-tilt model is unfortunately wrong.
\item The tilt-then-blur model is correct.
\item For natural images, the difference is less noticeable because many regions are smooth. For point sources, the difference can be significant.
\end{itemize}

\begin{figure*}[!]
\centering
\includegraphics[width=\linewidth]{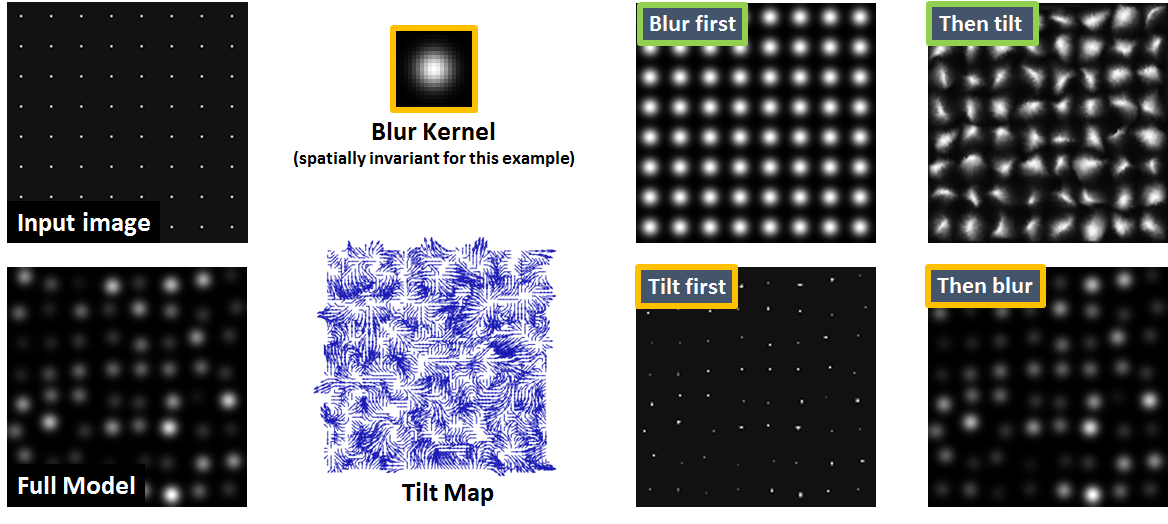}
\vspace{-4ex}
\caption{Comparing blur-then-tilt $\calT \circ \calB$ and tilt-then-blur $\calB \circ \calT$. Given a grid of point sources, a spatially invariant blur, and a dense tilt map, the result of $\calT \circ \calB$ shows a grid of destroyed blurs whereas the result of $\calB \circ \calT$ shows a grid of shifted blurs. We remark that $\calT \circ \calB$ is incorrect whereas $\calB \circ \calT$ is correct.}
\label{fig: Fig1 overview}
\end{figure*}

\section{Modeling Atmospheric Turbulence}
As a wave propagates through the atmospheric turbulence, the phase is distorted by the changing index of refraction. From the image formation point of view, if the clean image is $J(\vx)$ where $\vx \in \R^2$ specifies the coordinate in the 2D space, the distortion can be modeled by a spatially varying point spread function (PSF) $h_{\vx}(\vu)$. If the image $J(\vx)$ is discretized over a grid of $N$ pixels $\{\vx_1,\ldots,\vx_N\}$, the observed image is
\begin{equation}
I(\vx_i) = \sum_{j=1}^N h_{\vx_i}(\vu_j) J(\vu_j), \quad i = 1,\ldots,N.
\label{eq: I = h J}
\end{equation}
Here, $\vx_i$ specifies the output coordinate at which the distorted pixel should be located, and $\vu_j$ is a running index for the weighted average. \eref{eq: I = h J} is the general form of a spatially \emph{varying} convolution. In the special case where the PSF is spatially invarying, $h_{\vx_i}(\vu_j)$ can be written as $h(\vx_i-\vu_j)$ which is the usual convolution kernel.

For atmospheric turbulence, if the source is incoherent (e.g., passive imaging systems without any laser transmitters), the PSF is generated by taking the Fourier transform of the phase $\phi_{\vx}(\vrho)$ via \cite{Goodman_2005_FourierOptics, Goodman_2015_StatisticalOptics}
\begin{equation}
h_{\vx}(\vu) = |\calF(e^{-j2\pi \phi_{\vx}(\vrho)})|^2.
\label{eq: h fft}
\end{equation}
Here, the phase function $\phi_{\vx}(\vrho)$ is defined \emph{per pixel} at each coordinate $\vx$. The variable $\vrho \in \R^2$ denotes the phase coordinate, which is a 2D coordinate in the Fourier space.

The randomness of $h_{\vx}(\vu)$ comes from the randomness of the phase $\phi_{\vx}(\vrho)$ \cite{Tatarski_1967_Book}. $\phi_{\vx}(\vrho)$ is constructed by cropping and propagating the wave through a sequence of phase screens sampled from the Kolmogorov power spectral density \cite{Schmidt_2010_Book,Roggemann_2012_Simulator,Roggemann_1996_Book,Hardie_2017_OpEng}. This is a computationally expensive process, but recent work has alleviated the difficulty by modeling $\phi_{\vx}(\vrho)$ at the \emph{aperture} \cite{Chimitt_2020_OpEng, Mao_2021_ICCV}. Using the Zernike polynomials $\{Z_m(\vrho) \,|\, m = 1,\ldots,M\}$ as the basis representation, Noll \cite{Noll_1976_Zernike} states that we can write $\phi_{\vx}(\vrho)$ as
\begin{align}
\phi_{\vx}(\vrho) = \sum_{m=1}^M a_{\vx,m} Z_m(\vrho),
\end{align}
where $\{a_{\vx,m}\,|\,m=1,\ldots,M\}$ denote the $M$ Zernike coefficients at coordinate $\vx$. The Zernike coefficients are sampled from a zero-mean Gaussian random process
\begin{equation}
a_{\vx,m} \sim \calN(0, \mSigma),
\end{equation}
where the $(\vx,\vx',m,m')$th component of $\mSigma$ is $[\mSigma]_{\vx,\vx',m,m'} = \E[a_{\vx,m}a_{\vx',m'}]$. The joint expectation follows from Chanan \cite{Chanan_1992}, Takato and Yamaguchi \cite{Takato1995}, and Chimitt and Chan \cite{Chimitt_2020_OpEng}.

The representation of $\phi_{\vx}(\vrho)$ in the Zernike space gives us a way to decouple the \emph{shift} (ie, tilt) and the \emph{smoothing} (ie, blur). These two operations can be summarized as follows:
\begin{itemize}
\item Tilt $\calT$: The tilt is encoded by the first two Zernike bases via $\phi_{\vx}(\vrho) = \sum_{m=1}^2 a_{\vx,m}Z_m(\vrho)$;
\item Blur $\calB$: The blur is encoded by the remaining Zernike bases via $\phi_{\vx}(\vrho) = \sum_{m=3}^\infty a_{\vx,m}Z_m(\vrho)$.
\end{itemize}
However, how do we combine these two operations? There are a few options, based on the ordering of the tilt and the blur:
\begin{align}
\text{Blur-then-tilt}:\;\;\; &I(\vx) = [\calT \circ \calB](J(\vx)) = \calT(\calB(J(\vx))),\\
\text{Tilt-then-blur}:\;\;\; &I(\vx) = [\calB \circ \calT](J(\vx)) = \calB(\calT(J(\vx))),\\
\text{Full model}: \;\;\;    &I(\vx) = \calH(J(\vx)),
\end{align}
where ``$\circ$'' denotes the functional composition. For reference, the full turbulence model $I(\vx) = \calH(J(\vx))$ uses all the Zernike coefficients \emph{simultaneously} without decoupling them into tilts and blurs.

\section{Which One is Correct?}
The simplest way to compare the models is to run a numerical simulation and see which one is correct. Consider a grid of points as shown in \fref{fig: Fig1 overview}. We apply the three respective models $\calB \circ \calT$, $\calT \circ \calB$, and $\calH$ to the points and observe the resulting point spread functions (PSFs). The numerical experiment shows that $\calH = \calT \circ \calB$, and $\calT \circ \calB \not= \calB \circ \calT$. Let's discuss the reasons.

\subsection{Analysis from Matrices and Vectors}
We discretize the image so that the operations by $\calT$ and $\calB$ can be written in terms of matrices and vectors. Specifically, the operator $\calT$ can be written as a shifting matrix $\mT \in \R^{N \times N}$ with the $(i,j)$th entry being $[\mT]_{ij} = t_{\vx_i}(\vu_j)$, where $t_{\vx_i}(\vu_j) = 1$ if a pixel located at $\vu_j$ is relocated to coordinate $\vx_i$, and $t_{\vx_i}(\vu_j) = 0$ if otherwise. For example, if $J(\vx)$ is a 1D signal and $\mT$ shifts the signal by one pixel, then $\mT$ takes the form
\begin{equation}
\mT
=
\begin{bmatrix}
0 & 1 & 0 & 0 & \ldots & 0\\
0 & 0 & 1 & 0 & \ldots & 0\\
\vdots & \vdots & \vdots & \vdots & \ddots & \vdots \\
0 & 0 & 0 & 0 & \ldots & 0
\end{bmatrix}.
\label{eq: tilt example}
\end{equation}

The operator $\calB$ is a collection of tilt-free but spatially varying blurs. In the matrix notation, we can define a matrix $\mB$ where $[\mB]_{ij} = b_{\vx_i}(\vu_j)$ where $b_{\vx_i}$ is the tilt-free blur located at $\vx_i$. As before, $b_{\vx_i}$ is generated by the phase distortion at $\vx_i$ using high-order Zernike coefficients. The overall structure of the matrix $\mB$ is
\begin{equation}
\mB =
\begin{bmatrix}
b_{\vx_1}(\vu_1) & b_{\vx_1}(\vu_2) & \ldots & b_{\vx_1}(\vu_N)\\
b_{\vx_2}(\vu_1) & b_{\vx_2}(\vu_2) & \ldots & b_{\vx_2}(\vu_N)\\
\vdots           & \vdots           & \ddots & \vdots \\
b_{\vx_N}(\vu_1) & b_{\vx_N}(\vu_2) & \ldots & b_{\vx_N}(\vu_N)
\end{bmatrix}.
\end{equation}

At this point, it is easy to understand why $\calT\circ\calB$ and $\calB\circ\calT$ cannot be the same because matrices do not commute, i.e., $\mT\mB \not= \mB\mT$. In fact, if the tilt matrix is the one shown in \eref{eq: tilt example}, then $\mB\mT$ shifts $\mB$ to the left
\begin{align*}
\mB\mT =
\begin{bmatrix}
b_{\vx_1}(\vu_2) & \ldots & b_{\vx_1}(\vu_{N}) & 0\\
b_{\vx_2}(\vu_2) & \ldots & b_{\vx_2}(\vu_{N}) & 0\\
\vdots           & \vdots & \ddots             & \vdots \\
b_{\vx_N}(\vu_2) & \ldots & b_{\vx_N}(\vu_{N}) & 0
\end{bmatrix},
\end{align*}
whereas $\mT\mB$ shifts $\mB$ upwards
\begin{align*}
\mT\mB
=
\begin{bmatrix}
b_{\vx_2}(\vu_1) & b_{\vx_2}(\vu_2) & \ldots & b_{\vx_2}(\vu_N)\\
\vdots           & \vdots           & \ddots & \vdots \\
b_{\vx_N}(\vu_1) & b_{\vx_N}(\vu_2) & \ldots & b_{\vx_N}(\vu_N)\\
0                & 0                & 0      & 0
\end{bmatrix}.
\end{align*}

Now, the question is: which one is correct? Let us go back to \fref{fig: Fig1 overview}. Imagine that there is only one point source located at $\vx$. This point source is a delta function $J(\vu) = \delta(\vx-\vu)$ which will be distorted by the PSF $h(\vu)$ via $I(\vx) = \sum_j h(\vu_j)\delta(\vx-\vu_j)=h(\vx)$. Suppose that $h(\vx)$ is shifted by an amount $\vt$. Physically, the shift is caused by a phase offset in the exponential as can be seen in \eref{eq: h fft}. The tilt changes $I(\vx) = h(\vx)$ to $I(\vx) = h(\vx+\vt)$.

Using the tilt example shown in \eref{eq: tilt example}, the overall operator $\calH$ is to replace the blur $b_{\vx_i}(\vu_j)$ by $b_{\vx_i}(\vu_{j+1})$. Therefore, in terms of matrices and vectors, $\calH$ can be written as a matrix $\mH \in \R^{N \times N}$ where
\begin{equation}
\mH =
\begin{bmatrix}
b_{\vx_1}(\vu_2) & \ldots & b_{\vx_1}(\vu_{N}) & 0\\
b_{\vx_2}(\vu_2) & \ldots & b_{\vx_2}(\vu_{N}) & 0\\
\vdots           & \vdots & \ddots             & \vdots \\
b_{\vx_N}(\vu_2) & \ldots & b_{\vx_N}(\vu_{N}) & 0
\end{bmatrix} = \mB\mT.
\end{equation}
Notice that the shift occurs \emph{horizontally} in the matrix because we are shifting the blur. We are not relocating the output value to a different pixel location.

\subsection{Analysis from Geometry}
For readers who like a geometric explanation, here is the analysis. Assume that the blur is spatially invariant so that each blur can be written as
\begin{equation}
b_{\vx_i}(\vu) = g(\vx_i-\vu).
\end{equation}
Also, for notation simplicity, let us rewrite the tilt operator $\calT$ as a set of tilt vectors $\{\vt_i \;|\; i = 1,\ldots,N\}$. Each tilt vector $\vt_i$ is used to relocate pixel $J(\vx_i)$ to a new coordinate $\vx_i+\vt_i$:
\begin{equation*}
I_\calT(\vx_i+\vt_i) = J(\vx_i), \quad i = 1,\ldots,N.
\end{equation*}
This new notation can avoid the complication arising from the matrix-vector multiplication.

To clarify different combination of operations, we define the intermediate results $I_{\calB}$ and $I_{\calT}$ for blur and tilt only, and the final results $I_{\calB\circ\calT}$ and $I_{\calT\circ\calB}$ for the two operators.

\begin{figure}[h]
\centering
\includegraphics[width=\linewidth]{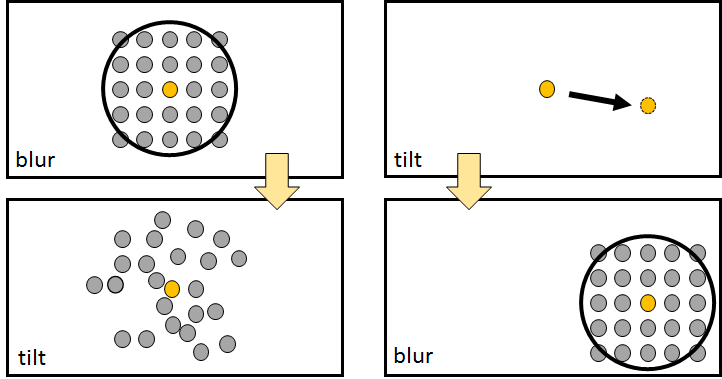}
\vspace{-4ex}
\caption{$\calT \circ \calB$ moves the center to a new location, whereas $\calB \circ \calT$ moves the points of the blur to different locations.}
\label{fig: Fig2 compare}
\end{figure}

\textbf{Blur-then-Tilt}. When a blur is applied to the clean image $J(\vu)$, the intermediate result $I_{\calB}(\vx)$ is
\begin{align*}
I_{\calB}(\vx_i)
&= \sum_{j=1}^N b_{\vx_i}(\vu_j)J(\vu_j) = \sum_{j=1}^N g(\vx_i-\vu_j)J(\vu_j).
\end{align*}
This is simply a blurred point source. Now consider the tilt. The tilt $\vt_i$ assigns $I_{\calB}(\vx_i)$ to a new pixel location $\vx_i+\vt_i$ of the final image $I_{\calT\circ\calB}$. That is,
\begin{equation*}
I_{\calT\circ\calB}(\vx_i + \vt_i) = I_{\calB}(\vx_i).
\end{equation*}
Since $\vt_1,\ldots,\vt_N$ is a dense field, $\calT$ will move \emph{every pixel} individually to a new location. As a result, the blur is not shifted uniformly to a new coordinate but each pixel is shifted to a different coordinate. Thus, the blur is \emph{destroyed}, as shown in \fref{fig: Fig2 compare}.

To prepare for a follow up discussion, we let $\vv_i = \vx_i + \vt_i$ and so $I_{\calT\circ\calB}(\vv_i) = I_{\calB}(\vv_i-\vt_i)$. Since $\vv_i$ is a dummy variable, $I_{\calT\circ\calB}$ becomes
\begin{equation}
I_{\calT\circ\calB}(\vx_i) = \sum_{j=1}^N g(\vx_i-\vt_i-\vu_j)J(\vu_j).
\label{eq: I TB}
\end{equation}

\textbf{Tilt-then-Blur}. When tilt is applied to the clean image $J(\vu)$, the intermediate result $I_{\calT}(\vx)$ is
\begin{equation}
I_{\calT}(\vu_j+\vt_j) = J(\vu_j).
\end{equation}
If now a blur is applied to $I_{\calT}$, the blur will simply move to the new location $\vu_j+\vt_j$. The shape of the blur is \emph{preserved}.

To complete the discussion. we let $\vv_j = \vu_j+\vt_j$. Then $I_{\calT}(\vv_j) = J(\vv_j-\vt_j)$, and
\begin{align*}
I_{\calB\circ\calT}(\vx_i)
&= \sum_{j=1}^N b_{\vx_i}(\vv_j) I_{\calT}(\vv_j)= \sum_{j=1}^N g(\vx_i-\vv_j) J(\vv_j-\vt_j).
\end{align*}
Replacing $\vu_j = \vv_j-\vt_j$, it follow that
\begin{equation}
I_{\calB\circ\calT}(\vx_i) = \sum_{j=1}^N g(\vx_i-\vt_j-\vu_j) J(\vu_j).
\label{eq: I BT}
\end{equation}

Comparing \eref{eq: I TB} and \eref{eq: I BT}, the difference between the two equations is that for $\calT\circ\calB$, the tilt is $\vt_i$ whereas for $\calB\circ\calT$ the tilt is $\vt_j$. In the case of $\calT \circ \calB$, the tilt $\vt_i$ is applied to the \emph{output}. That is, we blur the image (drawn as a circle in \fref{fig: Fig2 compare}) and move the output by $\vt_i$. Since each output pixel experiences a different $\vt_i$, the blur is destroyed. For $\calB \circ \calT$, although the situation is more complicated because there are $N$ tilts $\{\vt_j \;|\; j = 1\ldots,N\}$, they perturb the location of the \emph{input}. Therefore, while each $\vx_i$ sees $N$ tilts, the $N$ tilts are common for every $\vx_i$. The shape of the blur is thus preserved.

\section{Impact to Natural Images}
Although the two operators $\calT\circ\calB$ and $\calB\circ\calT$ are theoretically different, their impacts to real images are less so if the images do not contain any point sources.

To see why this is the case, rewrite \eref{eq: I TB} and \eref{eq: I BT} as
\begin{align*}
I_{\calT\circ\calB}(\vx_i)
&= \sum_{j=1}^N g(\vx_i-\vu_j)J(\vu_j-\vt_i),\\
I_{\calB\circ\calT}(\vx_i)
&= \sum_{j=1}^N g(\vx_i-\vu_j)J(\vu_j-\vt_j).
\end{align*}
Then, by approximating $J(\vu_j-\vt_i)$ and $J(\vu_j-\vt_j)$ to the first order, the pointwise difference between $I_{\calT\circ\calB}(\vx_i)$ and $I_{\calB\circ\calT}(\vx_i)$ can be evaluated as
\begin{align}
&I_{\calT\circ\calB}(\vx_i) - I_{\calB\circ\calT}(\vx_i) \notag\\
&\qquad= \sum_{j=1}^N g(\vx_i-\vu_j)\Big[J(\vu_j-\vt_i) - J(\vu_j-\vt_j)\Big] \notag\\
&\qquad\approx \sum_{j=1}^N \underset{\text{convolution}}{\underbrace{g(\vx_i-\vu_j)}} \;\;\; \underset{\text{distorted image gradient}}{\underbrace{
\underset{\text{image gradient}}{\underbrace{\nabla J(\vu_j)^T}} \;\;\;
\underset{\text{random tilt}}{\underbrace{(\vt_i-\vt_j)}}}}.
\end{align}

An intuitive argument here is that $\vt_i-\vt_j$ is the difference between two tilt vectors. Since each tilt is a zero-mean Gaussian random vector, the difference remains a zero-mean Gaussian random vector. Although they are not white Gaussian, they are nevertheless \emph{noise}. If there is a large ensemble average of these noise vectors, the result will be close to zero.

So, where does the average comes from? There is a convolution by $g(\vx-\vu)$. If the support of this blur kernel is large, then many of the noise vectors will be added and this will result in a small value. However, for an image with a large field of view, the relative size of the blur kernel $g$ is usually not big (at most $30 \times 30$ for a $256\times256$ image). So, there must be another source that makes the error small.

\begin{figure}[h]
\centering
\begin{tabular}{cc}
\includegraphics[width=0.44\linewidth]{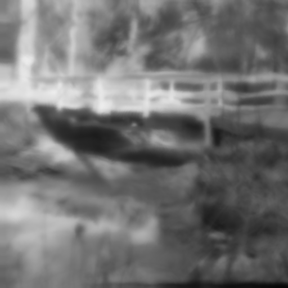}&
\includegraphics[width=0.44\linewidth]{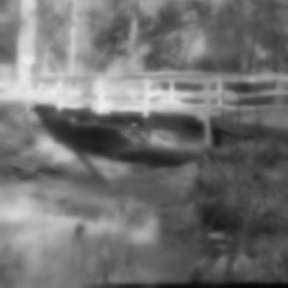}\\
(a) $\calT\circ\calB$ & (b) $\calB\circ\calT$
\end{tabular}
\caption{Simulated turbulence using $\calT\circ\calB$ and $\calB\circ\calT$. The optical parameters are as follows: Aperture diameter = 0.2034m, wavelength = 0.525$\mu$m, optical path = 7km, $C_n^2 = 5\times 10^{-6}$m$^{-2/3}$, focal length = 1.2m. }
\vspace{-2ex}
\label{fig: Fig3 boat}
\end{figure}

The main reason why natural images tend to show a less difference between $\calT\circ\calB$ and $\calB\circ\calT$ is that the image gradient $\nabla J(\vu_j)$ is typically \emph{sparse}. For most parts, the gradient is zero except for edges. (Textures are less of a problem because they will be smoothed by the blur.) When $\nabla J(\vu_j)$ is multiplied with the noise vector $\vt_i-\vt_j$, the result is an edge map with noise multiplied to every pixel. Convolving $\nabla J(\vu_j)^T(\vt_i-\vt_j)$ with a blur kernel $g$ will further smooth out the variations. \fref{fig: Fig3 boat} shows a typical example with some standard optical configurations. The difference is not noticeable.

\section{Recommendations}
This paper illustrated the validity of the tilt-then-blur model and issues of the blur-then-tilt model. The recommendation here is that when describing the imaging through turbulence problem, statements such as
\begin{center}
``\emph{... The image formation of atmospheric turbulence follows the equation $I(\vx) = \calT(\calB(J(\vx)))$ ... }''
\end{center}
\noindent should be avoided because they are wrong. Instead, it is more appropriate to state that the image formation model is $I(\vx) = \calB(\calT(J(\vx)))$ and comment that for natural images it can be approximated by $I(\vx) = \calT(\calB(J(\vx)))$.

From a simulation point of view, the computational complexity of implementing $\calT\circ\calB$ and $\calB\circ\calT$ are identical if one chooses the Zernike-based multi-aperture model \cite{Chimitt_2020_OpEng}. Since $\calB\circ\calT$ is \emph{the} correct model, there is no reason to go with an inferior model.

When solving inverse problems, however, there is a bit more freedom. One can choose to recover $\calT$ first, or one can choose to recover $\calB$ first, or simultaneously. Speaking of the author's own experience, recovering $\calB$ is less recommended because it is spatially varying. Therefore, even though the correct forward model is $I(\vx) = \calB(\calT(J(\vx)))$ so that the proper inversion is $J(\vx) = \calT^{-1}(\calB^{-1}(I(\vx)))$, it is often easier to estimate $\calT$ first. The residue error is usually not a problem when deep neural networks are used \cite{Zhang_2022_arXiv, Mao_2022_ECCV}.

In any case, the author hopes that the confusion between blur-then-tilt and tilt-then-blur is settled.

\bibliographystyle{IEEEbib}
\bibliography{refs}

\begin{thebibliography}{10}

\bibitem{Anantrasirichai_2013_TIP}
N.~Anantrasirichai, A.~Achim, N.~G. Kingsbury, and D.~R. Bull,
\newblock ``Atmospheric turbulence mitigation using complex wavelet-based
  fusion,''
\newblock {\em IEEE Transactions on Image Processing}, vol. 22, no. 6, pp.
  2398--2408, Jun. 2013.

\bibitem{Mao_2020_TCI}
Z.~Mao, N.~Chimitt, and S.~H. Chan,
\newblock ``Image reconstruction of static and dynamic scenes through
  anisoplanatic turbulence,''
\newblock {\em IEEE Transactions on Computational Imaging}, vol. 6, pp.
  1415--1428, Oct. 2020.

\bibitem{Li_2021_ICCV}
N.~Li, S.~Thapa, C.~Whyte, A.~W. Reed, S.~Jayasuriya, and J.~Ye,
\newblock ``Unsupervised non-rigid image distortion removal via grid
  deformation,''
\newblock in {\em Proceedings of the IEEE/CVF International Conference on
  Computer Vision (ICCV)}, October 2021, pp. 2522--2532.

\bibitem{Jin_2021_NatureMI}
D.~Jin, Y.~Chen, Y.~Lu, J.~Chen, P.~Wang, Z.~Liu, S.~Guo, and X.~Bai,
\newblock ``Neutralizing the impact of atmospheric turbulence on complex scene
  imaging via deep learning,''
\newblock {\em Nature Machine Intelligence}, vol. 3, pp. 876--884, 2021.

\bibitem{Feng_2022_TurbuGAN}
B.~Y. Feng, M.~Xie, and C.~A. Metzler,
\newblock ``{TurbuGAN}: An adversarial learning approach to spatially-varying
  multiframe blind deconvolution with applications to imaging through
  turbulence,'' 2022,
\newblock Available online: \url{https://arxiv.org/pdf/2203.06764.pdf}.
  Accessed 6/30/2022.

\bibitem{Goodman_2005_FourierOptics}
J.~W. Goodman,
\newblock {\em Introduction to Fourier Optics},
\newblock Roberts and Company, Englewood, Colorado, 3 edition, 2005.

\bibitem{Goodman_2015_StatisticalOptics}
J.~W. Goodman,
\newblock {\em Statistical Optics},
\newblock John Wiley and Sons Inc., Hoboken, New Jersey, 2 edition, 2015.

\bibitem{Roggemann_1996_Book}
M.~C. Roggemann and B.~M. Welsh,
\newblock {\em Imaging through Atmospheric Turbulence},
\newblock Laser \& Optical Science \& Technology. Taylor \& Francis, 1996.

\bibitem{Schmidt_2010_Book}
J.~D. Schmidt,
\newblock {\em Numerical simulation of optical wave propagation: With examples
  in {MATLAB}},
\newblock SPIE Press, Jan. 2010.

\bibitem{Tatarski_1967_Book}
V.~I. Tatarski,
\newblock {\em Wave Propagation in a Turbulent Medium},
\newblock New York: Dover Publications, 1961.

\bibitem{Fried_1965_Statistics}
D.~L. Fried,
\newblock ``Statistics of a geometric representation of wavefront distortion,''
\newblock {\em Journal of the Optical Society of America}, vol. 55, no. 11, pp.
  1427--1435, Nov. 1965.

\bibitem{Fried_1966_Optical}
D.~L. Fried,
\newblock ``Optical resolution through a randomly inhomogeneous medium for very
  long and very short exposures,''
\newblock {\em Journal of Optical Society of America}, vol. 56, no. 10, pp.
  1372--1379, 1966.

\bibitem{Shimizu_2008_CVPR}
M.~Shimizu, S.~Yoshimura, M.~Tanaka, and M.~Okutomi,
\newblock ``Super-resolution from image sequence under influence of hot-air
  optical turbulence,''
\newblock in {\em Proc. IEEE Intl' Conf. Computer Vision and Pattern
  Recognition (CVPR)}, 2008, pp. 1--8.

\bibitem{Leonard_2012_SPIE}
K.~R. Leonard, J.~Howe, and D.~E. Oxford,
\newblock ``Simulation of atmospheric turbulence effects and mitigation
  algorithms on stand-off automatic facial recognition,''
\newblock in {\em Proc. SPIE 8546, Optics and Photonics for Counterterrorism,
  Crime Fighting, and Defence VIII}, Oct. 2012, pp. 1--18.

\bibitem{Schwarzman_2017_ICCP}
A.~Schwartzman, M.~Alterman, R.~Zamir, and Y.~Y. Schechner,
\newblock ``Turbulence-indueced {2D} correlated image distortion,''
\newblock in {\em Proc. International Conference on Computational Photography},
  2017, pp. 1--12.

\bibitem{Milanfar_2013_TPAMI}
X.~Zhu and P.~Milanfar,
\newblock ``Removing atmospheric turbulence via space-invariant
  deconvolution,''
\newblock {\em IEEE Transactions on Pattern Analysis and Machine Intelligence},
  vol. 35, no. 1, pp. 157--170, Jan. 2013.

\bibitem{Gilles_2012_Inverse}
Y.~Mao and J.~Gilles,
\newblock ``Non rigid geometric distortions correction - application to
  atmospheric turbulence stabilization,''
\newblock {\em Inverse Problems and Imaging}, vol. 3, pp. 531--546, 2012.

\bibitem{Frakes_2001_ICASSP}
D.~H. Frakes, J.~W. Monaco, and M.~J.~T. Smith,
\newblock ``Suppression of atmospheric turbulence in video using an adaptive
  control grid interpolation approach,''
\newblock in {\em Proc. IEEE Intl' Conf. Acoustics, Speech, and Signal
  Processing (ICASSP)}, 2001, pp. 1881--1884.

\bibitem{Gepshtein_2004_EUSIP}
A.~Shteinman S.~Gepshtein and B.~Fishbain,
\newblock ``Restoration of atmospheric turbulent video containing real motion
  using rank filtering and elastic image registration,''
\newblock in {\em Proc. European Signal Processing Conference}, Sep. 2004.

\bibitem{Lou_2013_Inverse}
Y.~Lou, S.~Ha~Kang, S.~Soatto, and A.~Bertozzi,
\newblock ``Video stabilization of atmospheric turbulence distortion,''
\newblock {\em Inverse Problems and Imaging}, vol. 7, no. 3, pp. 839--861, Aug.
  2013.

\bibitem{Lau_2021_ATFaceGAN}
C.~P. Lau, H.~Souri, and R.~Chellappa,
\newblock ``Atfacegan: Single face semantic aware image restoration and
  recognition from atmospheric turbulence,''
\newblock {\em IEEE Transactions on Biometrics, Behavior, and Identity
  Science}, vol. 3, no. 2, pp. 240--251, Feb. 2021.

\bibitem{Nair_2021_ICIP}
N.~G. Nair and V.M. Patel,
\newblock ``Confidence guided network for atmospheric turbulence mitigation,''
\newblock in {\em Proc. IEEE Intl. Conf. Image Processing (ICIP)}, 2021, pp.
  1359--1363.

\bibitem{Yasarla_2022_CNN}
R.~Yasarla and V.~M. Patel,
\newblock ``{CNN-Based} restoration of a single face image degraded by
  atmospheric turbulence,''
\newblock {\em IEEE Transactions on Biometrics, Behavior, and Identity
  Science}, vol. 4, no. 2, pp. 222--233, 2022.

\bibitem{Lau_2017_Quasiconformal}
C.~P. Lau, Y.~H. Lai, and L.~M. Lui,
\newblock ``Restoration of atmospheric turbulence-distorted images via {RPCA}
  and quasiconformal maps,''
\newblock {\em Inverse Problems}, Mar. 2019.

\bibitem{Lau_2021_Subsampled}
C.~P. Lau and L.~M. Lui,
\newblock ``Subsampled turbulence removal network,''
\newblock {\em Mathematics, Computation and Geometry of Data}, vol. 1, no. 1,
  pp. 1--33, 2021.

\bibitem{Roggemann_2012_Simulator}
J.~P. Bos and M.~C. Roggemann,
\newblock ``Technique for simulating anisoplanatic image formation over long
  horizontal paths,''
\newblock {\em Optical Engineering}, vol. 51, no. 10, pp. 1 -- 9, 2012.

\bibitem{Hardie_2017_OpEng}
R.~C. Hardie, J.~D. Power, D.~A. LeMaster, D.~R. Droege, S.~Gladysz, and
  S.~Bose-Pillai,
\newblock ``Simulation of anisoplanatic imaging through optical turbulence
  using numerical wave propagation with new validation analysis,''
\newblock {\em Optical Engineering}, vol. 56, no. 7, pp. 1 -- 16, 2017.

\bibitem{Chimitt_2020_OpEng}
N.~Chimitt and S.~H. Chan,
\newblock ``Simulating anisoplanatic turbulence by sampling intermodal and
  spatially correlated {Zernike} coefficients,''
\newblock {\em Optical Engineering}, vol. 59, no. 8, pp. 1 -- 26, 2020.

\bibitem{Mao_2021_ICCV}
Z.~Mao, N.~Chimitt, and S.~H. Chan,
\newblock ``Accelerating atmospheric turbulence simulation via learned
  phase-to-space transform,''
\newblock in {\em Proc. IEEE/CVF International Conference on Computer Vision
  (ICCV)}, October 2021, pp. 14759--14768.

\bibitem{Noll_1976_Zernike}
R.~J. Noll,
\newblock ``{Zernike} polynomials and atmospheric turbulence,''
\newblock {\em Journal of Optical Society of America}, vol. 66, no. 3, pp.
  207--211, Mar. 1976.

\bibitem{Chanan_1992}
G.~A. Chanan,
\newblock ``Calculation of wave-front tilt correlations associated with
  atmospheric turbulence,''
\newblock {\em Journal of Optical Society of America A}, vol. 9, no. 2, pp.
  298--301, Feb. 1992.

\bibitem{Takato1995}
N.~Takato and I.~Yamaguchi,
\newblock ``Spatial correlation of {Zernike} phase-expansion coefficients for
  atmospheric turbulence with finite outer scale,''
\newblock {\em Journal of Optical Society of America A}, vol. 12, no. 5, pp.
  958--963, May 1995.

\bibitem{Zhang_2022_arXiv}
X.~Zhang, Z.~Mao, N.~Chimitt, and S.~H. Chan,
\newblock ``Imaging through the atmosphere using turbulence mitigation
  transformer,'' Available online: \url{https://arxiv.org/abs/2207.06465}.
  Accessed 8/7/2022.

\bibitem{Mao_2022_ECCV}
Z.~Mao, A.~Jaiswal, Z.~Wang, and S.~H. Chan,
\newblock ``Single frame atmospheric turbulence mitigation: A benchmark study
  and a new physics-inspired transformer model,''
\newblock in {\em Proc. European Conference on Computer Vision 2022},
\newblock Available online: \url{https://arxiv.org/abs/2207.10040}. Accessed
  8/7/2022.

\end{thebibliography}
\end{document}